\documentclass[12pt]{article}

\usepackage{amssymb,amsmath,bbm}
\usepackage{graphicx}
\usepackage{fullpage}

\newcommand{\quotient}[1]{_{\hskip-2pt\lower1pt\hbox{$/$}\lower2pt\hbox{\hskip-1pt$#1$}}}
\newcommand{\ii}{\text{i}} 
\newcommand{\bucon}{\widehat{\cC}}

  \usepackage{hypbmsec}
  \usepackage[sort&compress]{natbib}
  \bibpunct{[}{]}{,}{n}{}{;} 
  \usepackage[plainpages=false]{hyperref}
  \newcommand{\fref}[1]{\autoref{#1}}
  
  \newcommand{\sref}[1]{\autoref{#1}}
  \usepackage{twoopt}
  \newcommandtwoopt{\sectionhyp}[3][][]{\section[#1][#2]{#3}}
  \newcommandtwoopt{\subsectionhyp}[3][][]{\subsection[#1][#2]{#3}}

\newcommand{\cC}{\mathcal{C}}

\newcommand{\cN}{\mathcal{N}}
\newcommand{\cO}{\mathcal{O}}

\newcommand{\cS}{\mathcal{S}}

\renewcommand{\a}{\alpha}

\newcommand{\z}{\zeta}

\renewcommand{\l}{\lambda}
\newcommand{\m}{\mu}

\newcommand{\p}{\pi}

\newcommand{\ch}{\chi}

\renewcommand{\o}{\omega}

\newcommand{\IC}{\mathbb{C}}

\newcommand{\IP}{\mathbb{P}}

\newcommand{\IZ}{\mathbb{Z}}

\def\place#1#2#3{\vbox to0pt{\kern-\parskip\kern-7pt
                             \kern-#2truein\hbox{\kern#1truein #3}
                             \vss}\nointerlineskip}

\begin{document}

\begin{center}
{\Huge Quotients of the conifold in compact\\
Calabi-Yau threefolds, and new \\  \vspace{.1in}
topological transitions} \\
\vspace{.5in}
Rhys Davies\footnote{\it rhys.davies@physics.ox.ac.uk} \\
\vspace{.15in}
{\it
Rudolf Peierls Centre for Theoretical Physics, \\
University of Oxford \\
1 Keble Rd, Oxford \\
OX1 3NP, UK}
\end{center}
\vspace{.5in}

\abstract{
The moduli space of multiply-connected Calabi-Yau threefolds is shown
to contain codimension-one loci on which the corresponding variety develops
a certain type of hyperquotient singularity.  These have local descriptions
as discrete quotients of the conifold, and are referred to here as
hyperconifolds.  In many (or possibly all) cases such a singularity can be
resolved to yield a distinct compact Calabi-Yau manifold.  These considerations
therefore provide a method for embedding an interesting class of singularities
in compact Calabi-Yau varieties, and for constructing new Calabi-Yau manifolds.
It is unclear whether the transitions described can be realised in string theory.
}

\newpage

\section{Introduction}

Calabi-Yau threefolds have played a central role in string theory
compactifications since the seminal work~\cite{CHSW}.  A vast
number of these manifolds have now been constructed, the best-known
classes of which are the complete intersections in products of
projective spaces (CICYs), and hypersurfaces in weighted $\IP^4$
or more general toric fourfolds
\cite{CICYs1,Green:1986ck,Candelas:1989hd,Kreuzer:2000xy}.

In order to obtain semi-realistic heterotic compactifications, it is
usually necessary to consider multiply-connected Calabi-Yau manifolds,
on which discrete Wilson lines can be used to help break unwanted
gauge symmetries.  Two of the best-known examples are the heterotic
model on Yau's `three generation' manifold
\cite{Tian:1986ic,Greene:1986bm,Greene:1986jb}, with fundamental
group $\IZ_3$, and the models constructed on a manifold with
fundamental group $\IZ_3{\times}\IZ_3$
\cite{Braun:2004xv,Donagi:2004ub,Braun:2005ux,Braun:2005nv}.
Recent progress has been made on calculating the spectra of many
heterotic models on these types of spaces, opening up new realistic
model-building possibilities \cite{Braun:2009mb,Anderson:2009mh}.  Several
such multiply-connected manifolds were known in the early days of string
phenomenology \cite{CHSW,CICYs2,Strominger:1985it}, and recently a
number of new examples with various fundamental groups have been
constructed in \cite{SHN}.  Any such manifold necessarily arises as a
quotient of a simply-connected covering space by a freely-acting discrete
group.

Smooth Calabi-Yau manifolds are not the only ones relevant to string
theory.  The moduli spaces of families of smooth Calabi-Yau manifolds
have boundaries corresponding to singular varieties, and these are
moreover at a finite distance as measured by the moduli space metric
\cite{Candelas:1988di}.  Even more remarkable is that moduli spaces
of topologically distinct families can meet along such singular loci, and in
fact it has been speculated that the moduli space of all Calabi-Yau
threefolds is connected in this way~\cite{ReidFantasy}.  It was shown soon
after that this was true for almost all known examples, and suggested that
the associated string vacua may also be unified as a single physical moduli
space \cite{Green:1988bp,Green:1988wa}.  A series of beautiful
papers in the 90's established that type II string theories can indeed pass
smoothly through singular geometries, realising spacetime topology
change via so-called `flops' and `conifold transitions'
\cite{Aspinwall:1993nu,Strominger:1995cz,Greene:1995hu}.  Conifold
transitions can also be used as a tool for finding new Calabi-Yau manifolds,
as in \cite{Batyrev:2008rp,SHN}.

The most generic singularities which occur in threefolds are ordinary
double points, or nodes, which are usually referred to as conifold
singularities in the physics literature.  The purpose of this paper is to
point out that for multiply-connected threefolds, there are worse
singularities which are just as generic, in that they also occur on
codimension-one loci in moduli space.  Specifically, if the moduli are
chosen such that the (generically free) group action on the covering space
actually has fixed points, these turn out always to be singular
points, generically nodes.  The singularities on the quotient are therefore
quotients of the conifold, and, as we will see, have local descriptions as
toric varieties.  Standard techniques from toric geometry are therefore
utilised throughout; the reader unfamiliar with these ideas can consult
one of several reviews in the physics literature
\cite{Aspinwall:1993nu,Skarke:1998yk,Bouchard:2007ik}, or the textbook
\cite{Fulton}.

Quotients of hypersurface singularities were dubbed `hyperquotients' by
Reid.  We will refer to the particular examples considered herein
as `hyperconifolds', or sometimes $G$-hyperconifolds to be explicit about
the quotient group $G$.   Although the toric formalism allows us to find
local crepant resolutions (i.e. resolutions with trivial canonical bundle) of
these singularities in each case, the important question is whether these
preserve the Calabi-Yau conditions when embedded in the compact varieties
of interest.  In particular, the existence of a K\"ahler form is a global
question.  For most of the examples we can argue that a Calabi-Yau resolution
does indeed exist for all varieties containing the singularities of interest, and
furthermore we can calculate the Hodge numbers of such a resolution.  This
therefore gives a systematic way of constructing new Calabi-Yau manifolds
from known multiply-connected ones.  By analogy with conifold transitions,
the process by which we pass from the original smooth Calabi-Yau through
the singular variety to its resolution will be dubbed a `hyperconifold transition'.
Like a conifold transition, the Hodge numbers of the new manifold are different,
but unlike flops or conifold transitions, the fundamental group also changes.

Quotients of the conifold have been considered previously in the physics
literature, mostly in the context of D3-branes at singularities (e.g.
\cite{Uranga:1998vf,vonUnge,Hyun:2004yj,OhTatar,OhTatarRG}),
although the most-studied group actions in this context have fixed-point
sets of positive dimension.  The most simple example in this paper, the
$\IZ_2$-hyperconifold, has however appeared in numerous papers
(e.g. \cite{Beasley:1999uz,Feng:2000mi,Lu:2002rk,Benvenuti:2005qb}, and
recently in the context of heterotic theories with flux \cite{Carlevaro:2009jx}),
while the $\IZ_3$ case appears in an appendix in \cite{Aldazabal:2000sa}.
To the best of my knowledge hyperconifold singularities have not before
been explicitly embedded in compact varieties.  This paper gives a general
method to find compactifications of string/brane models based on these
singularities.

The layout of the paper is as follows.  In \sref{sec:Z5quintic} the
$\IZ_5$ quotient of the quintic is presented as an example of a compact
Calabi-Yau threefold which develops a hyperconifold singularity.  The
toric description of such singularities is also introduced here.
\sref{sec:hyperconifolds} contains the main mathematical result of
the paper.  It is demonstrated that if one starts with a family of threefolds
generically admitting a free group action, then specialises to a sub-family
for which the action instead develops a fixed point, then this point is
necessarily a singularity (generically a node).  The quotient variety
therefore develops a hyperconifold singularity; the toric descriptions of
these are given, and their topology described.  In \sref{sec:resolutions}
Calabi-Yau resolutions are shown to exist for many of these singular varieties,
and the Hodge numbers of these resolutions are calculated.  In
\sref{sec:stringtransitions} a few initial observations are made relating to
the possibility of hyperconifold transitions being realised in string theory.
\vskip10pt
The notation used throughout the paper is as follows:
\begin{itemize}
\item
$\widetilde X$ is a generic member of a family of smooth Calabi-Yau
threefolds which admit a free holomorphic action of the group $G$.
\item
$X$ is the (smooth, Calabi-Yau) quotient $\widetilde X/G$.
\item
$\widetilde X_0$ is a ($G$-invariant) deformation of $\widetilde X$
such that the action of $G$ is no longer free.
\item
$X_0$ is the singular quotient space $\widetilde X_0/G$.  This can be thought of as
living on the boundary of the moduli space of smooth manifolds $X$.
\item
$\widehat X$ will denote a crepant resolution of $X_0$, with projection
$\p:\widehat X \to X_0$.  We will denote by $E$ the exceptional set of
this resolution.
\end{itemize}

\newpage
\sectionhyp
[A $\IZ_5$ example]
[A Z5 example]
{A $\IZ_5$ example}
\label{sec:Z5quintic}

We will begin with a simple example to illustrate the idea.  Consider a
quintic hypersurface in $\IP^4$, and denote such a variety by
$\widetilde X$.  Take homogeneous coordinates $x_i$ for the ambient
space, with $i\in \IZ_5$, so that such a hypersurface is given by $f=0$,
where
\begin{equation}\label{eq:quintic}
f = \sum A_{ijklm}\, x_i\, x_j\, x_k\, x_l\, x_m
\end{equation}
If we denote by $g_5$ the generator of the cyclic group
$G \cong \IZ_5$, we can define an action of this group on the ambient
$\IP^4$ as follows:
$$
g_5:~ x_i \to \z^i x_i~~\text{where}~~\z = e^{2\p \ii/5}
$$
$\widetilde X$ will be invariant under this action if $A_{ijklm}$ is zero
except when  $i+j+k+l+m \equiv 0$ mod 5.  It is easy to see that for a general such
choice of these coefficients, the $\IZ_5$ action on $\widetilde X$ has no
fixed points, so the quotient variety, denoted $X$, is smooth.  For special
choices of complex structure though, the hypersurface given by $f=0$
\emph{will} contain fixed points, and it is this case which will interest us
here.

Consider the fixed point\footnote{The analysis is the same for any of the
five fixed points of the $\IZ_5$ action.} $[1,0,0,0,0] \in \IP^4$, and take
local affine coordinates $y_a = x_a/x_0,~a=1,2,3,4$ around this point.
Then the $\IZ_5$ action is given by
$$
y_a \to \z^a y_a
$$
and an invariant polynomial must locally be of the form
$$
f = \a_0 +  y_1\, y_4 - y_2\, y_3 + \text{higher-order terms}
$$
where $\a_0 := A_{00000}$ is one of the constant coefficients in
\eqref{eq:quintic} and we have chosen the coefficients of the quadratic
terms by rescaling the coordinates\footnote{The quadratic terms
correspond to some quadratic form $\eta$ on $\IC^4$.  Assuming that
$\eta$ is non-degenerate, it will always take the given form in appropriate
coordinates.  For general choices of coefficients in \eqref{eq:quintic},
$\eta$ will indeed by non-degenerate.}.  If we make the special choice
$\a_0 = 0$ (which corresponds to a codimension one locus in the
moduli space of invariant hypersurfaces), we obtain a variety
$\widetilde X_0$ on which the action is no longer free.

But now we see what turns out to be a general feature of this sort of
situation:  when $\a_0 = 0$ we actually have $f = df = 0$ at the
fixed point, meaning it is a node, or conifold, singularity on
$\widetilde X_0$.  This means that on its quotient $X_0$ we get a
particular type of hyperquotient singularity.  We will now study this
singularity by the methods of toric geometry.

\subsectionhyp
[The conifold and $\IZ_5$-hyperconifold as toric varieties]
[The conifold and Z5-hyperconifold as toric varieties]
{The conifold and $\IZ_5$-hyperconifold as toric varieties}
We will take the conifold $\cC$ to be described in $\IC^4$ by the
equation 
\begin{equation} \label{eq:conifold}
y_1 y_4 - y_2 y_3 = 0
\end{equation}
This is a toric variety whose fan consists of a single cone, spanned
by the vectors
\begin{equation} \label{eq:conifoldfan}
\begin{split}
v_1 = (1,0,0)~,~~ v_2 = (1,1,1) \\
v_3 = (1,1,0)~,~~ v_4 = (1,0,1)
\end{split}
\end{equation}
We can see that the four vertices lie on a hyperplane; this is equivalent
to the statement that the conifold is a non-compact Calabi-Yau variety.

\begin{figure}[!ht]
\begin{center}
\includegraphics[width=.25\textwidth]{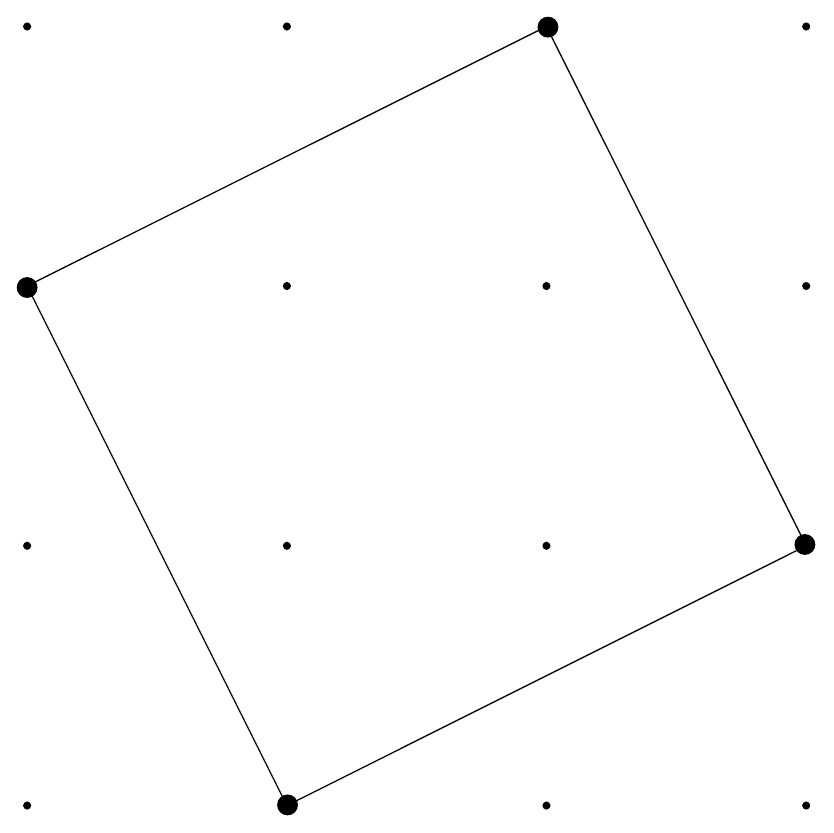}
\\
\parbox{.5\textwidth}
{\caption{\label{fig:Z5conifold}
\small The toric diagram for the $\IZ_5$-hyperconifold.}}
\end{center}
\end{figure}

We can also give homogeneous coordinates for the conifold, following
the prescription for toric varieties originally described by Cox:
$\cC = \big(\IC^4\setminus \cS\big)/\!\sim\,$, where the excluded set
is given by $\cS = \{z_1=z_2=0,\, (z_3,z_4) \neq (0,0)\}\cup
\{z_3=z_4=0,\, (z_1,z_2) \neq (0,0)\}$, and the equivalence relation
is
\begin{equation}\label{eq:homogconifold}
(z_1,z_2,z_3,z_4) \sim
(\l z_1, \l z_2, \l^{-1} z_3, \l^{-1}z_4)~\text{for}~ \l \in \IC^*
\end{equation}
The explicit isomorphism between this representation and the
hypersurface defined by \eqref{eq:conifold} is given by
\begin{equation}
y_1 = z_1 z_3~,~~ y_2 = z_1 z_4~,~~
  y_3 = z_2 z_3~,~~ y_4 = z_2 z_4
\end{equation}
The $\IZ_5$-hyperconifold singularity is obtained by imposing the
equivalence relation $(y_1,\, y_2,\, y_3,\, y_4) \sim
(\z\, y_1,\, \z^2\, y_2,\, \z^3\, y_3,\, \z^4\, y_4)$, 
where $\z = e^{2\p\ii/5}$.  Using the above equations we can express
this in terms of the $z$ coordinates as
$$
(z_1, z_2, z_3, z_4) \sim (z_1, \z^2 z_2, \z z_3, \z^2 z_4)
$$
This equivalence relation must imposed in addition to the earlier
one\footnote{Note that the power of $\z$ multiplying $z_1$ can
always be chosen to be trivial by simultaneously applying a
rescaling from~\eqref{eq:homogconifold}.}.  The resulting variety is
again a toric Calabi-Yau variety; the intersection of its fan with the
hyperplane on which the vertices lie is drawn in \fref{fig:Z5conifold}.
The singularity could easily be resolved by sub-dividing the fan, but
we will postpone a discussion of resolution of singularities until later.
First we want to prove that the example presented here is far from
unique.
\newpage
\section{Local hyperconifold singularities in general}
\label{sec:hyperconifolds}

The above discussion can be generalised to many other families
of multiply-connected Calabi-Yau threefolds.  To this end consider
a CY threefold $\widetilde X$ which, for appropriately chosen
complex structure, admits a free holomorphic action by some
discrete group $G$.  Then there exists a smooth quotient
$X = \widetilde X /G$, the deformations of which correspond to
$G$-invariant deformations of $\widetilde X$.  For simplicity I
will herein consider only the case in which $G$ is cyclic,
$G \cong \IZ_N$.  This is not a great restriction, since there seem
to be very few free actions of non-Abelian groups on Calabi-Yau
manifolds, and in any case, every non-Abelian group has Abelian
subgroups, to which the following discussion applies.

As we have seen for the $\IZ_5$-symmetric quintic, it may be that for
special choices of the complex structure of $X$ (generally on a
codimension-one locus in its moduli space) the action of $\IZ_N$ on
$\widetilde X$ will no longer be free. One might expect the resulting
singularities on $X$ to be simple orbifold singularities, locally of the form
$\IC^3/\IZ_N$.  In the case of the quintic though, we actually
obtained a quotient of the conifold.  We now demonstrate that this is
a general phenomenon.

\subsection{Analysis of fixed points}
\label{sec:singularproof}

Let $g_N$ denote the generator of $\IZ_N$, suppose that
$\widetilde X$ is locally determined by $k$ equations
$f_1  = \ldots = f_k = 0$ in $\IC^{k+3}$, on which some $\IZ_N$
action is given, and let $P_0 \in \IC^{k+3}$ be a fixed point of this
action.  Then we can choose local coordinates $x_1,\ldots,x_{k+3}$ at
$P_0$ such that the action of $g_N$ is given by $x_i \to \z^{q_i} x_i$,
where $\z = e^{2\p\ii/N}$ and $q_i \in \{ 0,\ldots, N-1 \}$.  Let $I$
be the set of fixed points of this action, and order the coordinates such
that $I$ is given locally by $x_{\dim I+1}=\ldots= x_{k+3}=0$.  This
is equivalent to $q_1 = \ldots = q_{\dim I} = 0$ and $q_i \neq 0$ for
$i > \dim I$.

By taking linear combinations of the polynomials if necessary, we can
assume that $g_N\cdot f_a = \z^{Q_a} f_a$.  What we mean by this
is that $f_a(g_N\cdot P) = \z^{Q_a} f_a(P)$ for $P \in \IC^{k+3}$.
This  immediately implies that if $Q_a \neq 0$, then we must have
$f_a|_I \equiv 0$.  But since by assumption $\widetilde X$ does not
generically intersect $I$, at least $\dim I+1$ of the polynomials must be
non-trivial when restricted to $I$, so that they have no common zeros.
We conclude that at least $\dim I +1$ of the polynomials must be
invariant under the group action.

Now suppose that we choose special polynomials such that the
corresponding variety $\widetilde X_0$ intersects $I$ at a point, and
identify this point with $P_0$ above:
$I \cap \widetilde X_0 = \{P_0\}$.  The expansion of an invariant
polynomial $f_a$ (i.e. $Q_a = 0$) around $P_0$ is then
$$
f_a = \sum_{i=1}^{\dim I} C_{a,i}\, x_i + \cO(x^2)
$$
Now we can see what goes wrong.  At $P_0$ we have
$$
\frac{\partial f_a}{\partial x_i} = \Bigg\{ \begin{array}{lcr}
 C_{a,i}&,& i \leq \dim I \\[5pt]
 0 &,& i > \dim I \end{array}
$$ 
so the matrix $\partial f_a/\partial x_i$, for $f_a$ ranging over invariant
polynomials, has maximal rank $\dim I$.  But since, as argued above,
there are at least $\dim I + 1$ invariant polynomials, at the point $P_0$
we get $f_a = 0$ for all $a$ and
$$
df_1\wedge \ldots \wedge df_{\dim I + 1} = 0~~~\text{and hence}~~~
df_1\wedge \ldots \wedge df_k = 0
$$
So the variety $\widetilde X_0$ is singular at this point, and in fact
generically it will have a node, or conifold singularity.  This means
that the quotient variety $X_0$ develops a worse local singularity: a
quotient of the conifold by a $\IZ_N$ action fixing only the singular point.
This is what we will now call a $\IZ_N$-hyperconifold.

It should be noted that there is no reason for any other singularities to
occur on $X_0$, and indeed it can be checked in specific cases
that only one singular point develops.

\subsection{The hyperconifolds torically}\label{sec:quotients}

\begin{figure}[!h]
\begin{center}
\framebox[6in]{\parbox{6in}{
\begin{center}
\includegraphics[width=.2\textwidth]{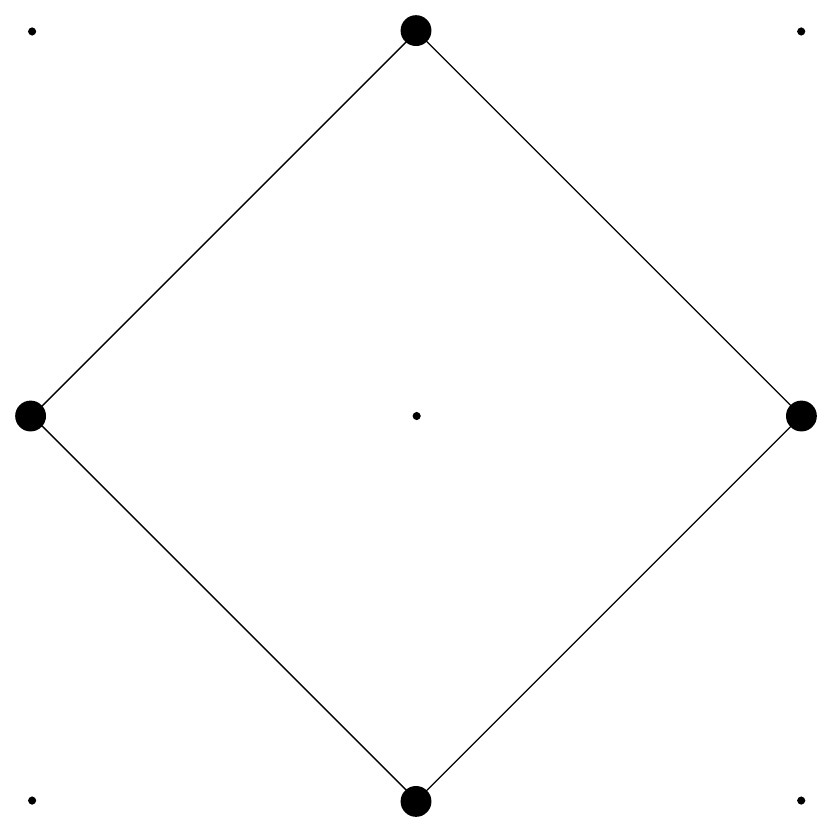}
\hskip85pt
\raisebox{-20pt}{\includegraphics[width=.2\textwidth]{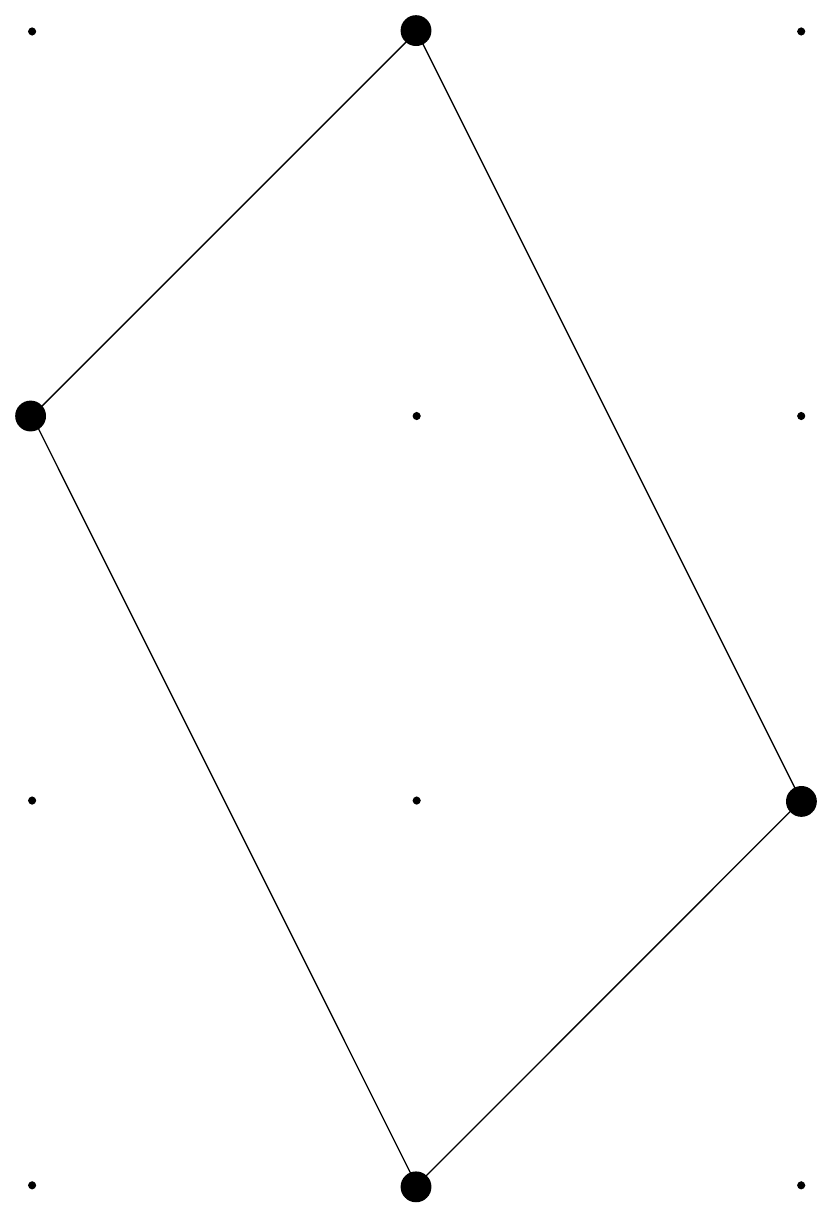}}\\[30pt]
\hskip23pt
\includegraphics[width=.15\textwidth]{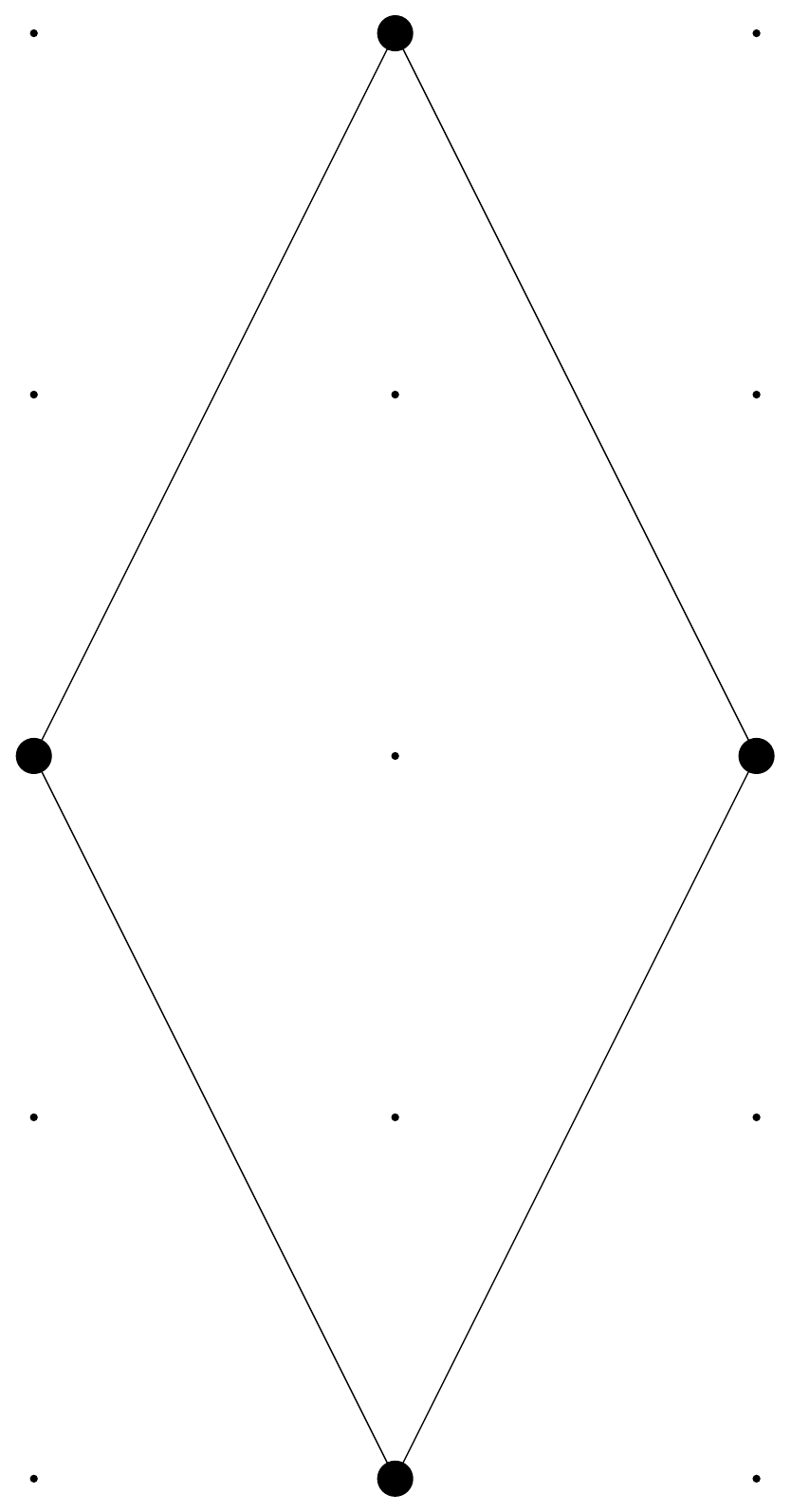}
\hskip85pt
\raisebox{8pt}{\includegraphics[width=.25\textwidth]{Z5conifold.pdf}}\\[35pt]
\hskip30pt
\includegraphics[width=.15\textwidth]{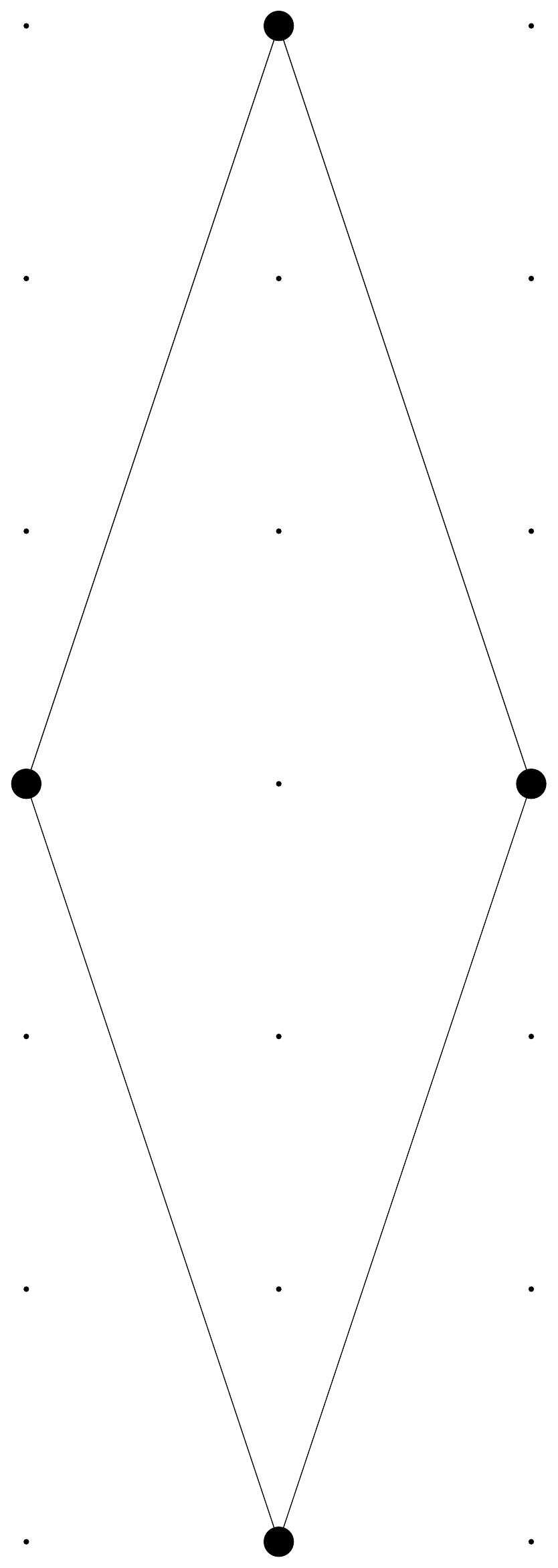}
\hskip70pt
\raisebox{30pt}{\includegraphics[width=.3\textwidth]{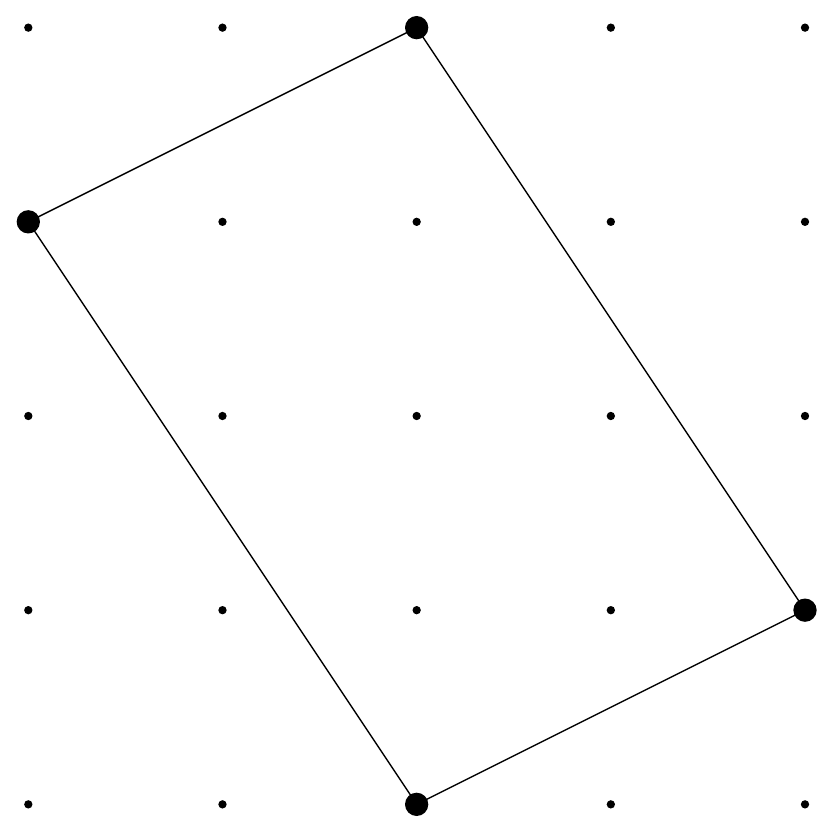}}
\end{center}
}}
\parbox{6in}{\caption{\small  \label{fig:hyperconifolds}
The toric diagrams for the $\IZ_N$-hyperconifolds,
where $N = 2,3,4,5,6,8$.
}}
\end{center}
\end{figure}

\begin{figure}[!h]
\begin{center}
\framebox[6in]{\parbox{6in}{
\begin{center}
\includegraphics[width=.4\textwidth]{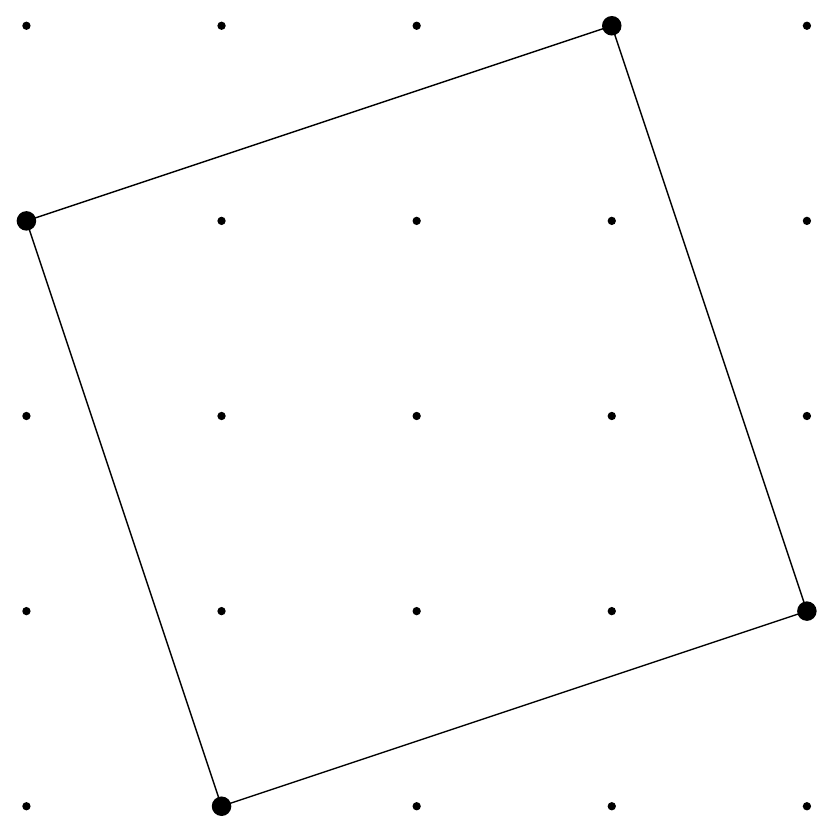} \\[60pt]
\includegraphics[width=.4\textwidth]{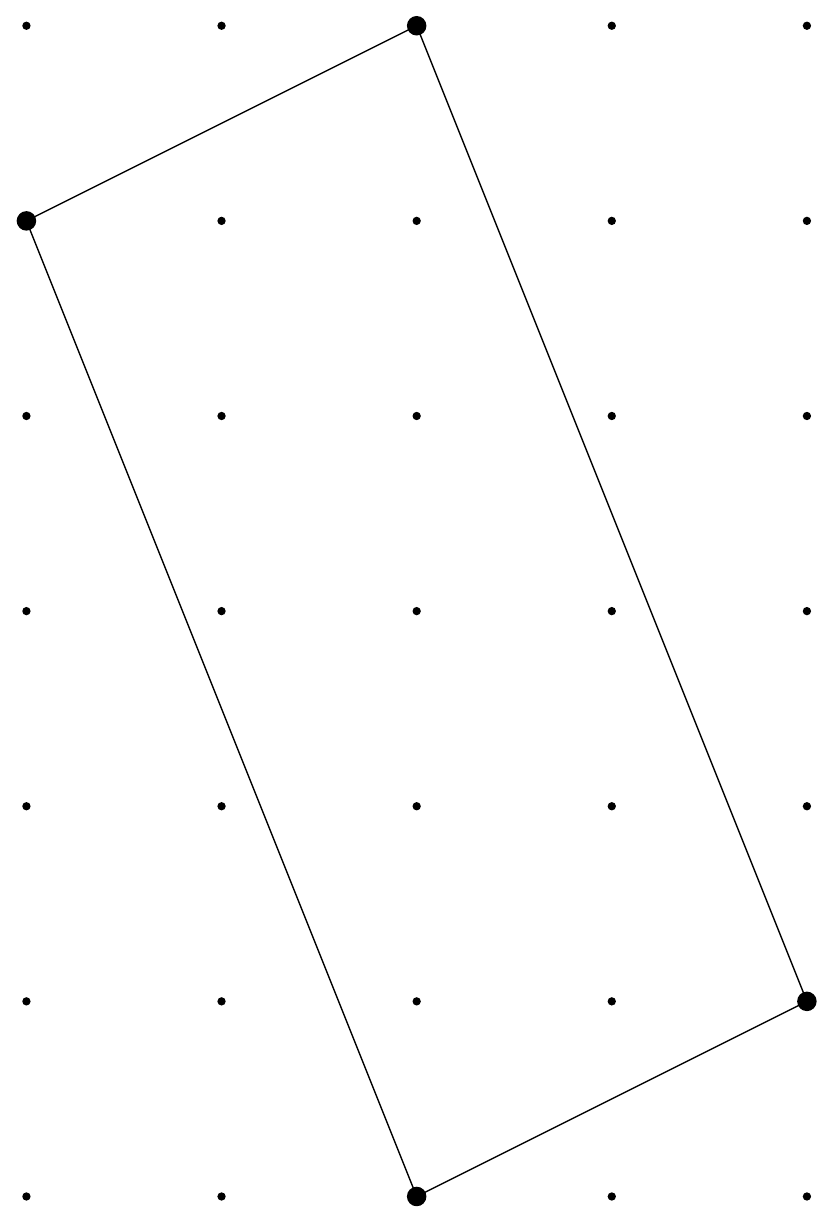}
\end{center}
}}
\parbox{6in}{\caption{\small  \label{fig:hyperconifolds2}
The toric diagrams for the $\IZ_{10}$- and $\IZ_{12}$-hyperconifold
singularities.
}}
\end{center}
\end{figure}

We now want to give explicit descriptions of the types of singularities
whose existence in compact Calabi-Yau varieties we demonstrated
above.  There are known Calabi-Yau threefolds with fundamental group
$\IZ_N$ for $N = 2,3,4,5,6,7,8,10,12$, and all cases except $N=7$
occur as quotients of CICYs
\cite{Strominger:1985it,SHN,GrossPopescuI,BCD}.  For these we can
perform analyses similar to that presented earlier for the $\IZ_5$
quotient of the quintic, and obtain singular varieties containing isolated
hyperconifold singularities.  Each of these has a local toric description,
which will be presented below.  Since these are all toric Calabi-Yau
varieties, the vectors generating the one-dimensional cones of their fans
lie on a hyperplane; \fref{fig:hyperconifolds} and
\fref{fig:hyperconifolds2} at the end of this section are collections of
diagrams showing the intersection of the fans with this hyperplane.  It
should be noted that from the diagrams it is obvious that each singularity
admits at least one toric crepant resolution.  However we are only
interested in those which give Calabi-Yau resolutions of the compact
variety in which the singularity resides.  Determining whether such a
resolution exists requires more work, which we defer to
\sref{sec:resolutions}.
\vskip15pt
\noindent
\emph{$\IZ_2$ quotient}\\
Note that, as discussed above, the only point on the conifold fixed
by the group actions we are considering will be the singular point itself.
As such, there is only a single possible action of
$\IZ_2$:
\begin{equation*}
(y_1, y_2, y_3, y_4) \to (-y_1, -y_2, -y_3, -y_4)
\end{equation*}
In terms of the homogeneous coordinates this gives the additional
equivalence relation
\begin{equation*}
(z_1, z_2, z_3, z_4) \sim (z_1, z_2,  -z_3, -z_4)
\end{equation*}
The resulting singularity is one which \emph{has} appeared in the
physics literature, as mentioned earlier.  The difference here is that we
have given a prescription for embedding this singularity in a compact
Calabi-Yau variety, in such a way that it admits both a smooth
deformation and, as we will see later, a resolution.
\newpage
\noindent
\emph{$\IZ_3$ quotient}\\
Similarly to the $\IZ_2$ case, there is only a single action of $\IZ_3$
on  the conifold with an isolated fixed point:
\begin{equation*}
(y_1,\, y_2,\, y_3,\, y_4) \to (\z\, y_1,\, \z\, y_2,\, \z^2\, y_3,\, \z^2\, y_4)
\end{equation*}
where $\z = e^{2\p \ii/3}$.  In terms of the homogeneous
coordinates this leads to
\begin{equation*}
(z_1,\, z_2,\, z_3,\, z_4) \sim (z_1,\, \z\, z_2,\, \z\, z_3,\, \z\, z_4)
\end{equation*}
\vskip15pt
\noindent
\emph{$\IZ_4$ quotient}\\
The group $\IZ_4$ has a $\IZ_2$ subgroup which must also act non-trivially on each
coordinate $y_a$, so again there is a unique action consistent with this:
\begin{equation*}
(y_1,\, y_2,\, y_3,\, y_4) \to (\ii\, y_1,\, \ii\, y_2,\, -\ii\, y_3,\, -\ii\, y_4)
\end{equation*}
In terms of the homogeneous coordinates this implies
\begin{equation*}
(z_1,\, z_2,\, z_3,\, z_4) \sim (z_1,\, - z_2,\, \ii\, z_3,\, \ii\, z_4)
\end{equation*}
\vskip15pt
\noindent
\emph{$\IZ_5$ quotient}\\
This is the first case where there are two actions of the group on the
conifold which fix only the origin.  This is true for $\IZ_5$ and several
of the larger cyclic groups discussed below, but in each case only one
of the actions actually occurs in known examples.  For $\IZ_5$ it is
\begin{equation}
(y_1,\, y_2,\, y_3,\, y_4) \to
(\z\, y_1,\, \z^2\, y_2,\, \z^3\, y_3,\, \z^4\, y_4)
\end{equation}
where $\z = e^{2\p \ii/5}$.  We have already seen this in our original
example of the quintic.  In terms of the homogeneous coordinates the
new equivalence relation is
\begin{equation*}
(z_1,\, z_2,\, z_3,\, z_4) \sim
(z_1,\, \z^2\, z_2,\, \z\, z_3,\, \z^2\, z_4)
\end{equation*}
\vskip15pt
\noindent
\emph{$\IZ_6$ quotient}\\
For $\IZ_6$ we can once again find the action by general arguments.
If we require all elements of the group to act with only a single fixed
point, there is only one possibility:
\begin{equation*}
(y_1,\, y_2,\, y_3,\, y_4) \to
(\z\, y_1,\, \z\, y_2,\, \z^5\, y_3,\, \z^5\, y_4)
\end{equation*}
where $\z = e^{\p \ii/3}$.  The identification on the homogeneous
coordinates is therefore
\begin{equation*}
(z_1,\, z_2,\, z_3,\, z_4) \sim (z_1,\, \z^4\, z_2,\, \z\, z_3,\, \z\, z_4)
\end{equation*}
\newpage
\noindent
\emph{$\IZ_8$ quotient}\\
As in the $\IZ_5$ case, there are multiple actions of $\IZ_8$ on
the conifold which fix only the origin, but only one is realised in the
present context.  The only free $\IZ_8$ actions I know on compact
Calabi-Yau threefolds are the one described in
\cite{Strominger:1985it} and those related to it by conifold
transitions~\cite{GrossPopescuI,BorisovHua}.  These can be
deformed to obtain a local conifold singularity with the following
quotient group action
\begin{equation*}
(y_1,\, y_2,\, y_3,\, y_4) \to
(\z\, y_1,\, \z^3\, y_2,\, \z^5\, y_3,\, \z^7\, y_4)
\end{equation*}
where $\z = \exp{\p \ii/4}$.  The equivalence relation on the
homogeneous coordinates is therefore
\begin{equation*}
(z_1,\, z_2,\, z_3,\, z_4) \sim
(z_1,\, \z^4\, z_2,\, \z\, z_3,\, \z^3\, z_4)
\end{equation*}
\vskip15pt
\noindent
\emph{$\IZ_{10}$ quotient}\\
Several free actions of $\IZ_{10}$ on Calabi-Yau manifolds were
described in \cite{SHN}.  If we allow one of these to develop a
fixed point, the resulting action on the conifold is
\begin{equation*}
(y_1,\, y_2,\, y_3,\, y_4) \to
(\z\, y_1,\, \z^3\, y_2,\, \z^7\, y_3,\, \z^9\, y_4)
\end{equation*}
where $\z = \exp{\p \ii/5}$.  The corresponding equivalence relation
on the homogeneous coordinates is
\begin{equation*}
(z_1,\, z_2,\, z_3,\, z_4) \sim
(z_1,\, \z^6\, z_2,\, \z\, z_3,\, \z^3\, z_4)
\end{equation*}
\vskip15pt
\noindent
\emph{$\IZ_{12}$ quotient}\\
The largest cyclic group known to act freely on a Calabi-Yau manifold
is $\IZ_{12}$, and this was discovered only recently \cite{BCD}.
The resulting action on the conifold is
\begin{equation*}
(y_1,\, y_2,\, y_3,\, y_4) \to
(\z\, y_1,\, \z^5\, y_2,\, \z^7\, y_3,\, \z^{11}\, y_4)
\end{equation*}
where $\z = \exp{\p \ii/6}$.  The corresponding equivalence relation
on the homogeneous coordinates is
\begin{equation*}
(z_1,\, z_2,\, z_3,\, z_4) \sim
(z_1,\, \z^6\, z_2,\, \z\, z_3,\, \z^5\, z_4)
\end{equation*}

\newpage
\newpage
%

\subsection{Topology of the singularities}

Topologically, the conifold is a cone over $S^3{\times}S^2$.
It would be nice to relate the group actions described herein to
this topology.  Evslin and Kuperstein have provided a convenient
parametrisation of the base of the conifold for just this sort of
purpose~\cite{Evslin:2007ux}, which I will use here:

Parametrise the conifold as the set of degenerate $2{\times}2$
complex matrices
\begin{equation}
W = \left(\begin{array}{cc}
y_1 & y_2 \\
y_3 & y_4 \end{array} \right)
~,~~~~ \det W = 0
\end{equation}
and identify the base with the subset satisfying
$\text{Tr} (W^\dagger W) = 1$.  Now identify $S^3$ with the
underlying topological space of the group $SU(2)$, and $S^2$
with the space of unit two-vectors modulo phases.  Then we
map the point $(X, v) \in S^3{\times}S^2$ to
\begin{equation}\label{eq:conifoldtriv}
W = X\, v\, v^\dagger
\end{equation}
This is shown to be a homeomorphism in~\cite{Evslin:2007ux}.
The actions of $\IZ_2, \IZ_3, \IZ_4, \IZ_6$, described in
\sref{sec:quotients}, are all realised in this description by
\begin{equation}
W \to \left(\begin{array}{ll}
\z & 0 \\
0 & \z^{-1}\end{array}\!\!\!\!\right) W
\end{equation}
where $\z = e^{2\p\ii/N}$, $N=2,3,4,6$ respectively.  We see from
\eqref{eq:conifoldtriv} that this can be considered as an action purely
on the $S^3$ factor of the base, the quotient by which is the lens
space $L(N,1)$.  Topologically then, the singularity is locally a cone
over $L(N,1){\times}S^2$.  In fact these same spaces were
considered in~\cite{Halmagyi:2003mm}.

The more complicated cases of $\IZ_5, \IZ_8, \IZ_{10}$ and
$\IZ_{12}$ quotients don't have such a straightforward topological
description, but could be analysed along the lines of \cite{Evslin:2007ux}.

\section{Global resolutions} \label{sec:resolutions}

In the preceding section we have described the local structure of the
hyperconifolds using toric geometry; now we want to address the
question of their resolution.  Certainly if we consider each case as a
non-compact variety, they can all easily be resolved using toric
methods.  The interesting question is whether or not the compact
varieties containing these singularities can be resolved to yield new
Calabi-Yau manifolds.

\subsection{Blowing up a node}

It will be useful to first consider blowing up an ordinary node, and
only then turn to its quotients.  Again we take the conifold $\cC$ to
be given in $\IC^4$ by \eqref{eq:conifold}:
$$
y_1 y_4 - y_2 y_3 = 0
$$
The singularity lies at the origin, and we can resolve it by blowing up
this point.  To do so we introduce a $\IP^3$ with homogeneous
coordinates $(t_1, t_2, t_3, t_4)$, and consider the equations
$y_i t_j - y_j t_i = 0$ in $\IC^4{\times}\,\IP^3$.  This has the effect
of setting $(t_1, t_2, t_3, t_4) \propto (y_1,y_2,y_3,y_4)$ when at
least one $y_i$ is non-zero, but leaving the $t$'s completely
undetermined at the origin.  In this way we `blow up' a single point
to an entire copy of $\IP^3$, and have a natural projection map
$\p$ which blows it down again.  The blow up of the conifold is then
defined to be the closure of the pre-image of its smooth points:
$$
\bucon = \overline{\p^{-1}(\cC\backslash\{\mathbf{0}\})}
$$
Therefore $\bucon$ is isomorphic to $\cC$ away from the
node, but the node itself is replaced by the surface in $\IP^3$ given
by
$$
t_1 t_4 - t_2 t_3 = 0
$$
which is in fact just $\IP^1{\times}\,\IP^1$.  This is called the
\emph{exceptional divisor} of the blow-up, and we will denote it by
$E$.  Another important piece of information is the normal bundle
$\cN_{E | \bucon}$ to $E$ inside $\bucon$.  If
$\cO(n,m)$ denotes the line bundle which restricts to the $n^\text{th}$
(resp. $m^\text{th}$) power of the hyperplane bundle on the first
(resp. second) $\IP^1$, then in this case the normal bundle is
$\cO(-1,-1)$.  This can be verified by taking an affine cover and writing
down transition functions, but the toric formalism, to which we turn
shortly, will let us see this much more easily.  In any case, with this
information we can demonstrate that $\widehat\cC$ is \emph{not} Calabi-Yau.
To see why, recall the adjunction formula for the canonical bundle of the
hypersurface $E$ in terms of that of $\bucon$
\begin{equation} \label{eq:adjunction}
\o_E = \o_{\bucon}\Big|_{E}\! \otimes \cN_{E|\bucon}
\end{equation}
Therefore if $\o_{\bucon}$ were trivial, we would have
$\o_E \cong \cN_{E|\bucon} \cong \cO(-1,-1)$, but it is a well-known fact
that $\o_E = \cO(-2,-2)$, so we conclude that $\bucon$ has a non-trivial
canonical bundle.  This is why the blow up of a node does not generally
feature in discussions of Calabi-Yau manifolds.  We will see soon why it
becomes relevant once we want to consider quotients.

The final important point is that the blowing up procedure automatically
gives us another projective variety (i.e. a K\"ahler manifold if it is
smooth), since the blow up is embedded in the product of the original
space with a projective space.
\vspace{-10pt}
\subsectionhyp
[The toric picture, and the $\IZ_2$-hyperconifold]
[The toric picture, and the Z2-hyperconifold]
{The toric picture, and the $\IZ_2$-hyperconifold}

\begin{figure}[ht]
\begin{center}
\includegraphics[width=.4\textwidth]{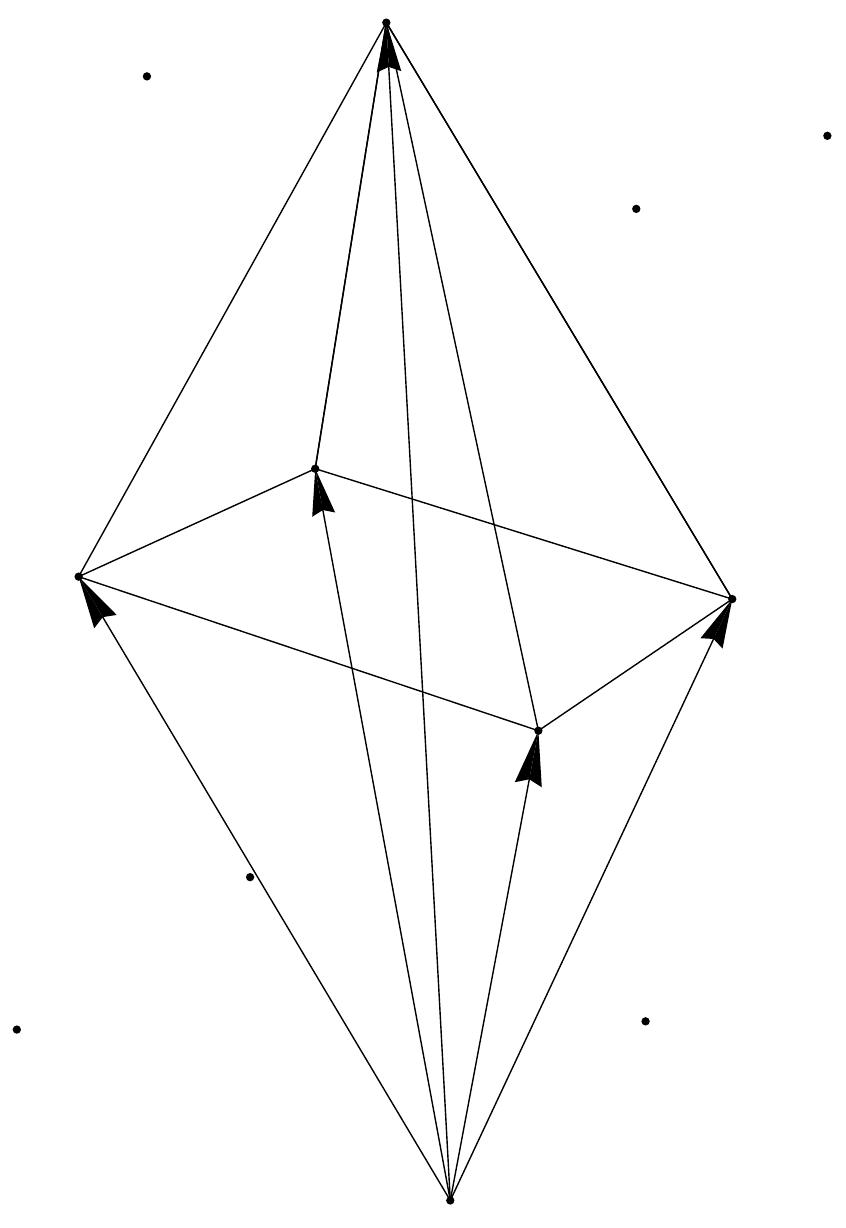}
\hskip60pt
\raisebox{30pt}{\includegraphics[width=.4\textwidth]{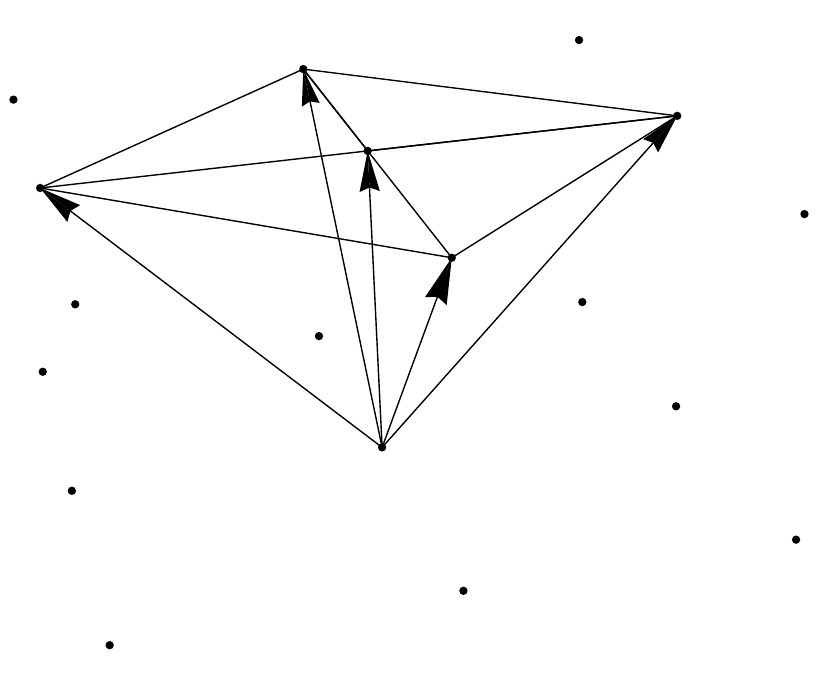}}
\vskip1pt
\place{1.95}{1.65}{$v_1$}
\place{1}{2.6}{$v_2$}
\place{2.55}{2.1}{$v_3$}
\place{.25}{2.15}{$v_4$}
\place{1.2}{3.9}{$v_5$}
\place{5.18}{1.9}{$v_1$}
\place{4.4}{2.6}{$v_2$}
\place{5.88}{2.35}{$v_3$}
\place{3.6}{2.15}{$v_4$}
\place{4.83}{2.4}{$v_5$}
\parbox{.8\textwidth}
{\caption{\label{fig:blownupconifold}
\small The fans for the blow-up of the conifold at its singular point,
and its $\IZ_2$ quotient.}}
\end{center}
\end{figure}

We can also blow up the node on the conifold using toric geometry.
Recall that the fan for $\cC$ consists of a single cone spanned by the
four vectors given in \eqref{eq:conifoldfan}, plus its faces.  To this set
of vectors we want to add $v_5 = v_1 + v_2 = v_3 + v_4$, and
sub-divide the fan accordingly.  The result is shown in
\fref{fig:blownupconifold}.
We now have five homogeneous coordinates, and two independent
rescalings (we won't explicitly describe the set to be removed before
taking the quotient -- this can be read from the fan).
\vspace{-3pt}
\begin{equation} \label{eq:homogbucon}
(z_1,z_2,z_3,z_4,z_5) \sim
(\l\, z_1,\l\, z_2,\m\, z_3,\m\, z_4, \l^{-1}\m^{-1}z_5)
\quad \l,\m \in \IC^*
\end{equation}
From this data we can easily see that $(z_1,z_2)$ parametrise a
$\IP^1$, as do $(z_3,z_4)$, and $z_5$ is a coordinate on the fibre of
$\cO(-1,-1)$.  When $z_5 \neq 0$, we can choose $\m = \l^{-1}\, z_5$,
and we obtain the isomorphism to $\cC\backslash\{\mathbf{0}\}$.
The remaining points are on the toric divisor given by $z_5=0$, and it
is clear that this is isomorphic to $\IP^1{\times}\, \IP^1$.  So this
toric variety is indeed the blow up of $\cC$ at the origin.  The toric
formalism makes it clear that the resolution is not crepant, since the
new vector does not lie on the same hyperplane as the others.

We now turn our attention to the $\IZ_2$-hyperconifold,
which was described in \sref{sec:quotients}, and which we will
denote by $\cC_2$.  In this case the blow up of the singular point is
obtained by adding a vector which lies on the same hyperplane as
the first four, meaning that the resulting resolution $\widehat \cC_2$
is also Calabi-Yau.  The relations are now
$2v_5 = v_1 + v_2 = v_3 + v_4$, and the five vectors can be
taken to be
\vspace{-5pt}
\begin{eqnarray*}
v_1 = (1,-1,0)~,& v_2 = (1,1,0) \\
v_3 = (1,0,-1)~,& v_4 = (1,0,1) \\
v_5 =& \hskip-40pt (1,0,0)
\end{eqnarray*}
The resulting equivalence relations on the homogeneous coordinates
are
\begin{equation}
(w_1,w_2,w_3,w_4,w_5) \sim
(\l\, w_1,\l\, w_2,\m\, w_3,\m\, w_4, \l^{-2}\m^{-2}w_5)
\quad \l,\m \in \IC^*
\end{equation}
\begin{figure}[t]
\begin{center}
\includegraphics[width=.3\textwidth]{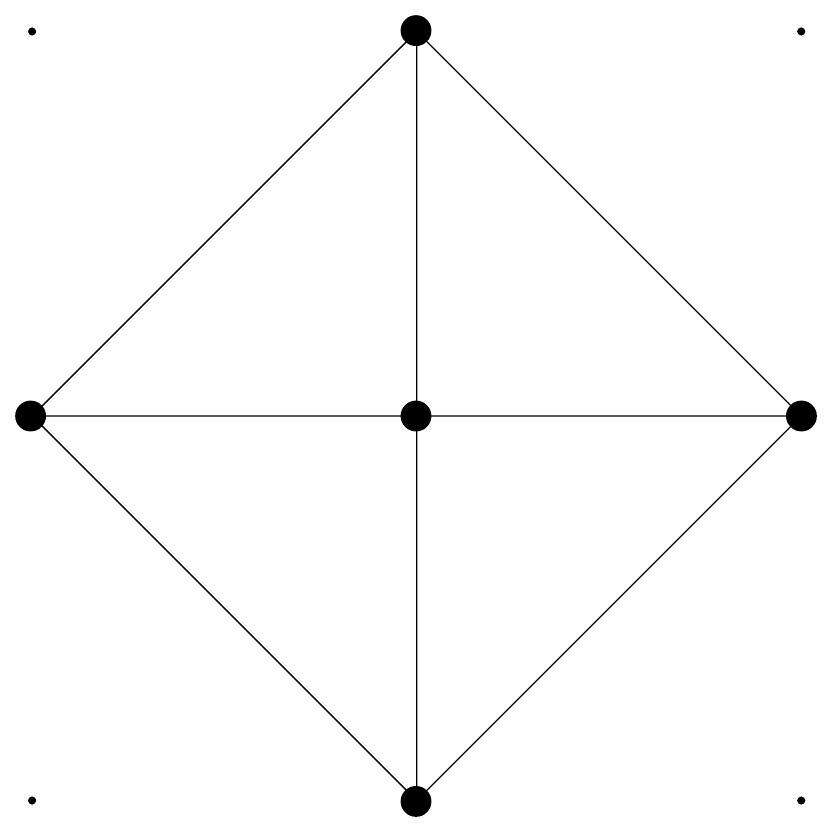}
\vskip1pt
\parbox{.8\textwidth}
{\caption{\label{fig:Z2blownupconifold}
\small The toric diagram for the resolution of the $\IZ_2$-hyperconifold.}}
\end{center}
\end{figure}
\vskip-5pt
\noindent This is very similar to \eqref{eq:homogbucon}, but in this case $w_5$
is seen to be a coordinate on $\cO(-2,-2)$ rather than $\cO(-1,-1)$.  As
such, the adjunction formula \eqref{eq:adjunction} says that the canonical
bundle of $\widehat\cC_2$ restricts to be trivial on the exceptional
divisor, consistent with $\widehat\cC_2$ being Calabi-Yau.

There is another nice way to think about the resolution of $\cC_2$.  We
begin by noticing that the blown-up conifold $\widehat\cC$ is actually a
ramified double cover of $\widehat\cC_2$, with the explicit map being
given by
\begin{equation}\label{eq:doublecover}
w_i = z_i ~~\text{for}~~ i=1,2,3,4 \quad\quad w_5 = (z_5)^2
\end{equation}
This deserves some elaboration.  It is clear from
\eqref{eq:doublecover} that the map is two-to-one everywhere
except along the exceptional divisor given by $z_5 = 0$.  In fact it
can be thought of as an identification $z_5 \sim -z_5$ on the fibres
of $\cO(-1,-1)$ over $\IP^1{\times}\,\IP^1$.  Since the fixed
point set of the involution $z_5\to-z_5$ is of complex codimension
one, taking the quotient actually does not introduce any singularity
(which is clear here since $\widehat\cC_2$ is manifestly smooth).
So we can think of the resolution of $\cC_2$ in two ways: either we
blow up the singular point of $\cC_2$, or we blow up the node on the
covering space, and then take the $\IZ_2$ quotient.

Note that the procedure described above is completely local (we
blew up a point), and therefore can be performed inside any compact
Calabi-Yau variety $X_0$ in which the singularity $\cC_2$ occurs, to yield
a new compact Calabi-Yau manifold.  This should be compared to the
small resolution of the conifold, which involves blowing up a
sub-variety which extends `to infinity' in $\cC$ (in fact the variety
given by $y_1 = y_2 = 0$), so that the existence of the Calabi-Yau
resolution depends on the global structure.
 
\vspace{-10pt}
\subsectionhyp
[The $\IZ_{2M}$-hyperconifolds]
[The Z2M-hyperconifolds]
{The $\IZ_{2M}$-hyperconifolds}

Having demonstrated the existence of crepant projective resolutions
(i.e. Calabi-Yau resolutions) for Calabi-Yau varieties containing the
singularity $\cC_2$, we can easily do the same for the quotients of the
conifold by all cyclic groups of even order.  This is achieved by breaking
the process down into several steps\footnote{The following argument is
partly due to Bal\'azs Szendr\H{o}i.}.

The unique $\IZ_2$ subgroup of $\IZ_{2M}$ fixes exactly the
singular point, and we can blow this up by adding the ray through
the point $v_5 = \frac 14 (v_1 + v_2 + v_3 + v_4)$ and sub-dividing
the fan accordingly.  Alternatively we can think of this as first taking
the quotient by $\IZ_2 \subset \IZ_{2M}$, blowing up the resulting
$\cC_2$ singularity, and then taking the quotient by the induced action of
$\IZ_{2M}/\IZ_2 \cong \IZ_M$.  Either way, we obtain a variety with
only $\IZ_M$ orbifold singularities.  There are then two cases:
\begin{itemize}
\item
If $M$ is odd, it turns out that
there is a unique way to further sub-divide the fan to obtain a
smooth variety.  In \cite{Chen} it is shown that for a projective
threefold with only orbifold singularities, one obtains a global projective
crepant resolution by choosing an appropriate crepant resolution
on each affine patch.  If there is a unique choice for each, we therefore
automatically obtain the projective resolution, so we are done.

\item
If $M$ is even, then $\IZ_M$ contains a unique $\IZ_2$ subgroup,
and the fixed point set of this subgroup is a pair of disjoint curves which
are toric orbits (this follows from inspecting the diagrams case-by-case).
These are given by two-cones, spanned by $v_i, v_j$, and in each case
the vector $\frac 12 (v_i + v_j)$ is integral, so can be added to the fan
to blow up the fixed curve (in fact this is just the well-known resolution
of the $A_1$ surface singularity, fibred over the curves).  We iterate this
process until we are left with $\IZ_{M'}$ orbifold singularities for
$M'$ odd, and the fan then has a unique smooth subdivision.
\end{itemize}
Note that I am not claiming that the resolutions obtained are the unique
K\"ahler ones.  Several of the hyperconifolds admit multiple resolutions
differing by flops, and it is possible that more than one of these corresponds
to a projective resolution.

\begin{figure}[p]
\begin{center}
\framebox[6in]{\parbox{6in}{
\begin{center}
\includegraphics[width=.31\textwidth]{Z12conifold.pdf}
\hskip75pt
\includegraphics[width=.31\textwidth]{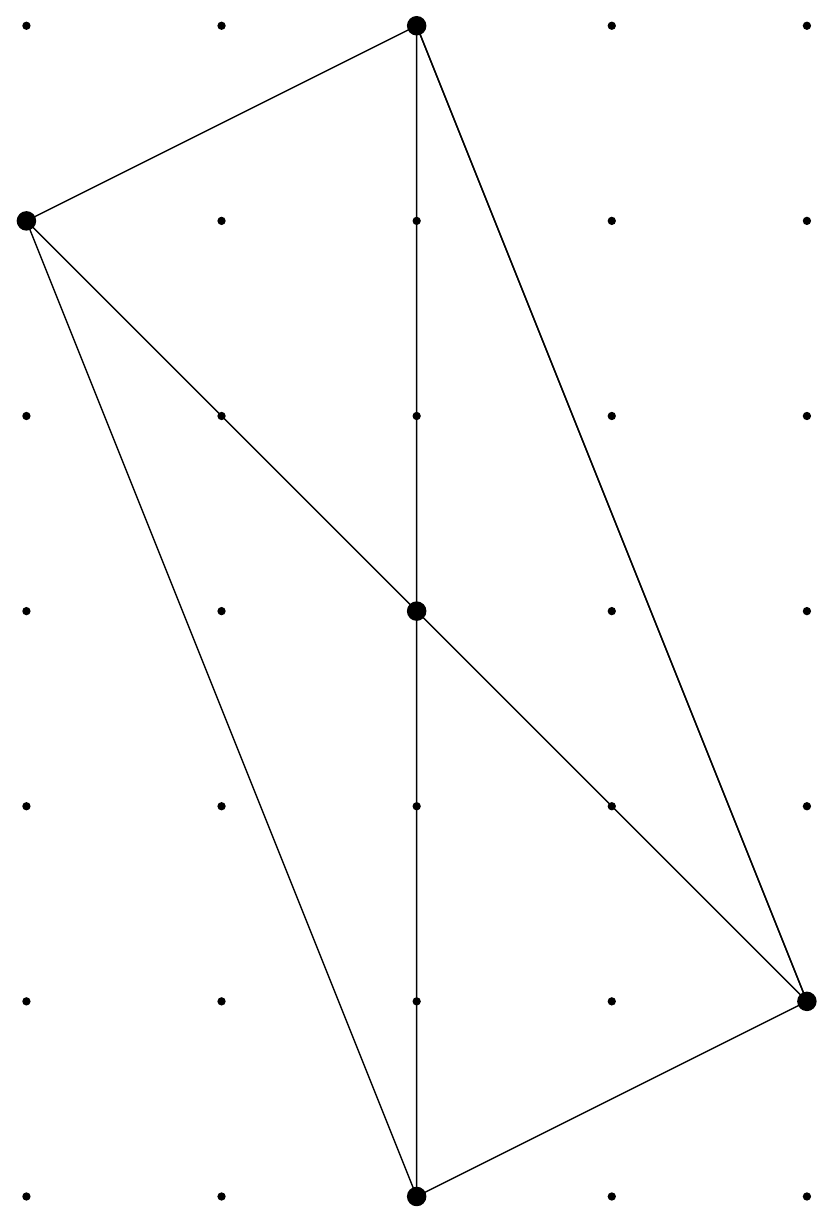}\\[70pt]
\includegraphics[width=.31\textwidth]{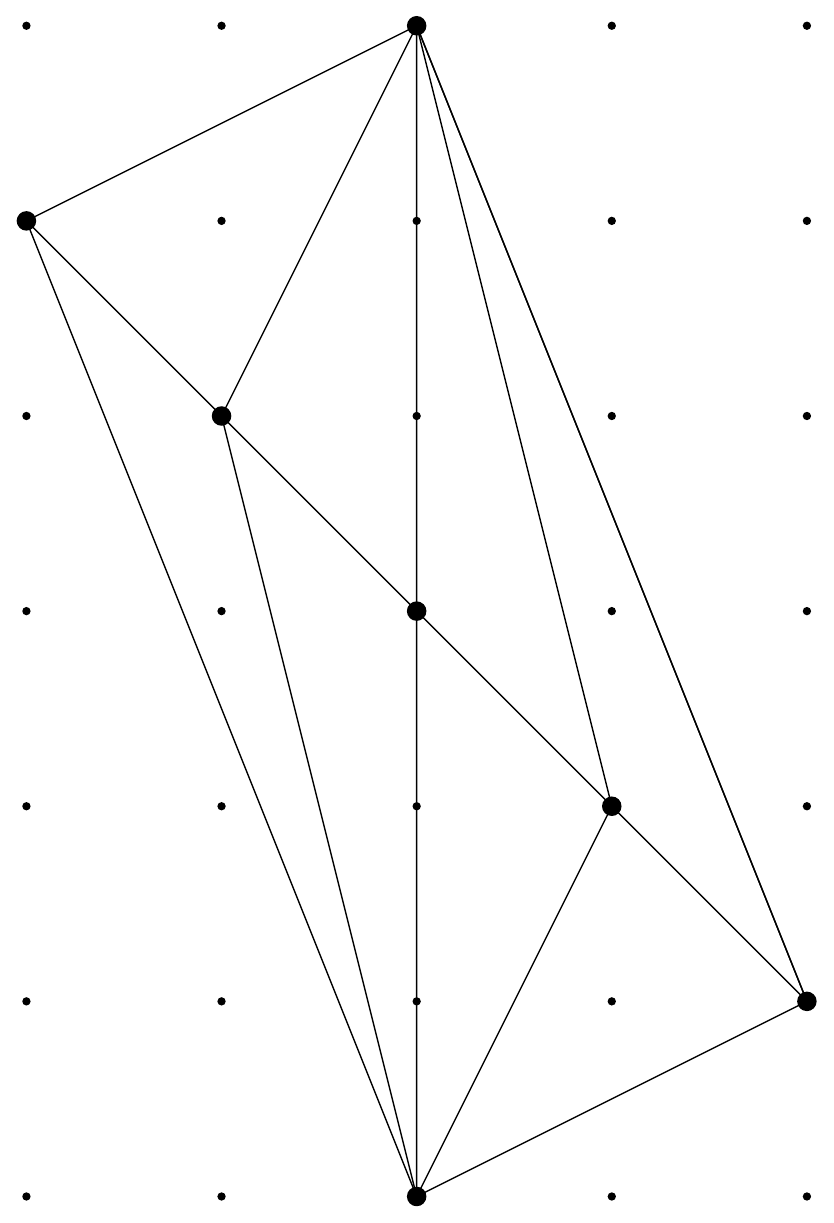}
\hskip75pt
\includegraphics[width=.31\textwidth]{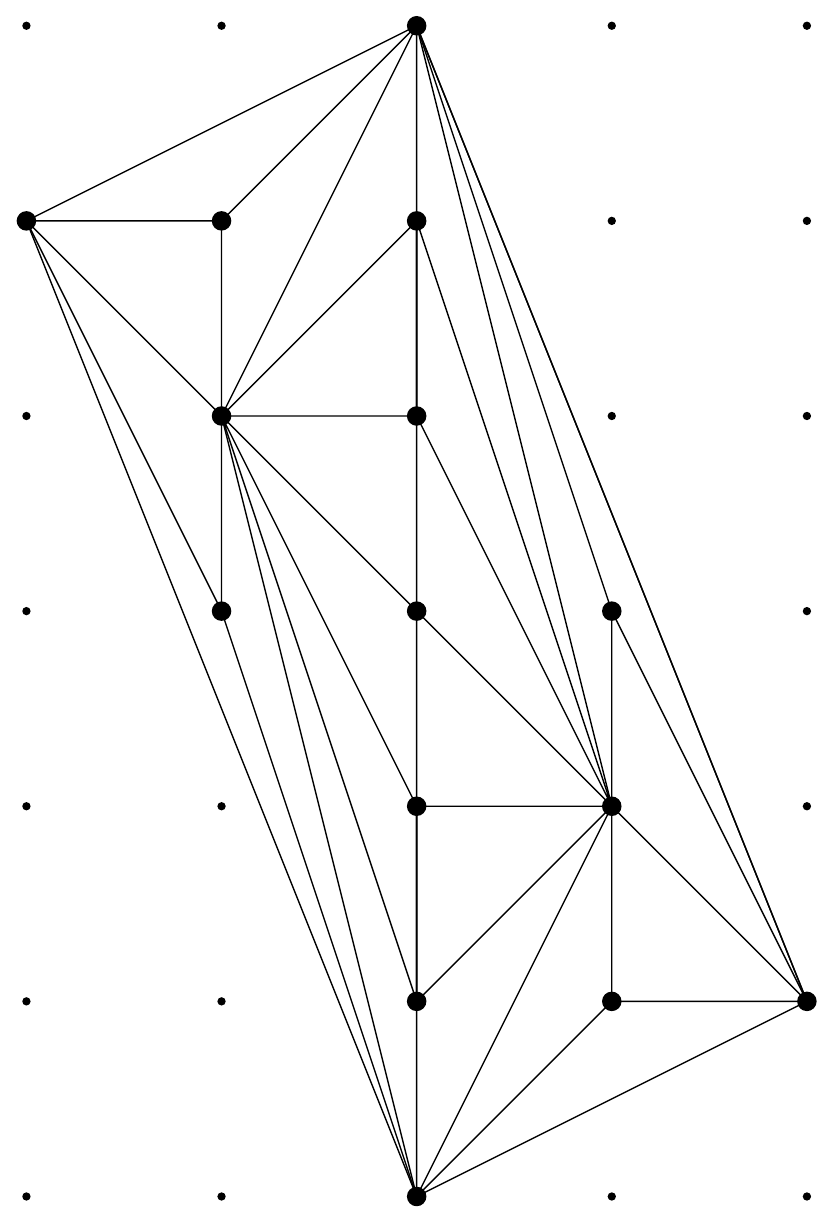}
\end{center}
}}
\parbox{6in}{\caption{\small  \label{fig:Z12resolution}
The three steps involved in resolving the $\IZ_{12}$-hyperconifold
singularity.  First we blow up the singular point, then the two curves fixed
by $\IZ_2 \subset \IZ_6$, and finally we perform the unique maximal
subdivision of the resulting fan.
}}
\end{center}
\end{figure}

The preceding prescription is easy to understand in particular cases,
as we will now illustrate with the complicated $\IZ_{12}$-hyperconifold.
We begin with the fan in \fref{fig:hyperconifolds2}, and blow up the
singular point, which adds a ray through the geometric centre of the
top-dimensional cone, and divides it into four (see
\fref{fig:Z12resolution}).  The result is a fan for a variety containing a
chain of four genus zero curves meeting in points.  Two of these are
curves of $\IC_2/\IZ_2$ orbifold singularities, and the other two of
$\IC_2/\IZ_3$ singularities.  The four points of intersection are locally
$\IC_3/\IZ_6$ orbifold singularities. We can blow up the (disjoint)
$\IZ_2$ curves by bisecting the corresponding two-cones and
sub-dividing the fan accordingly.  This leaves us with eight
top-dimensional simplicial cones, each of which has a unique crepant
sub-division, giving us the final smooth, crepant, K\"ahler resolution
of the singularity.

\begin{figure}[p]
\begin{center}
\framebox[6in]{\parbox{6in}{
\begin{center}
\raisebox{30pt}{\includegraphics[width=.15\textwidth]{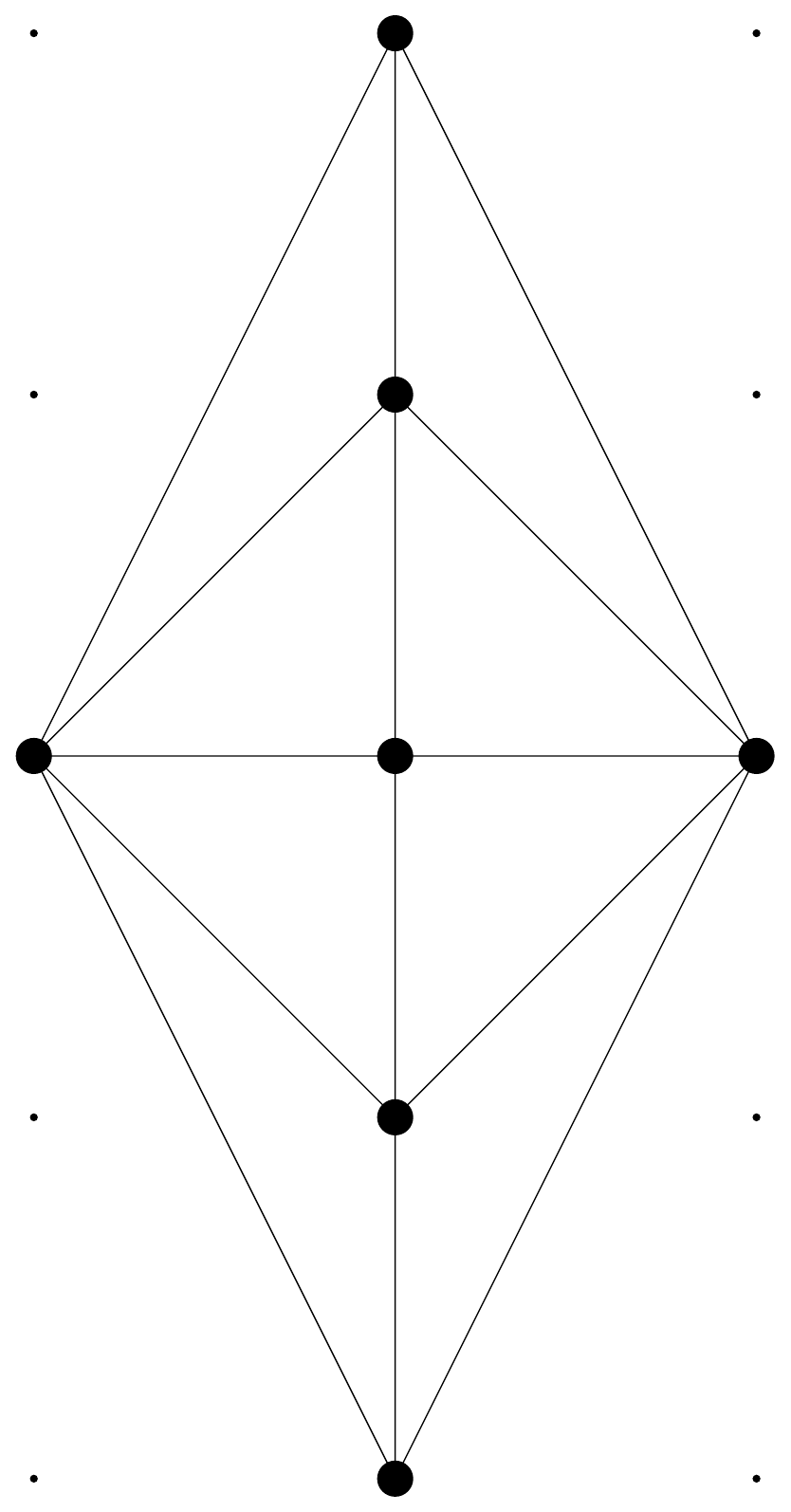}}
\hskip145pt
\includegraphics[width=.15\textwidth]{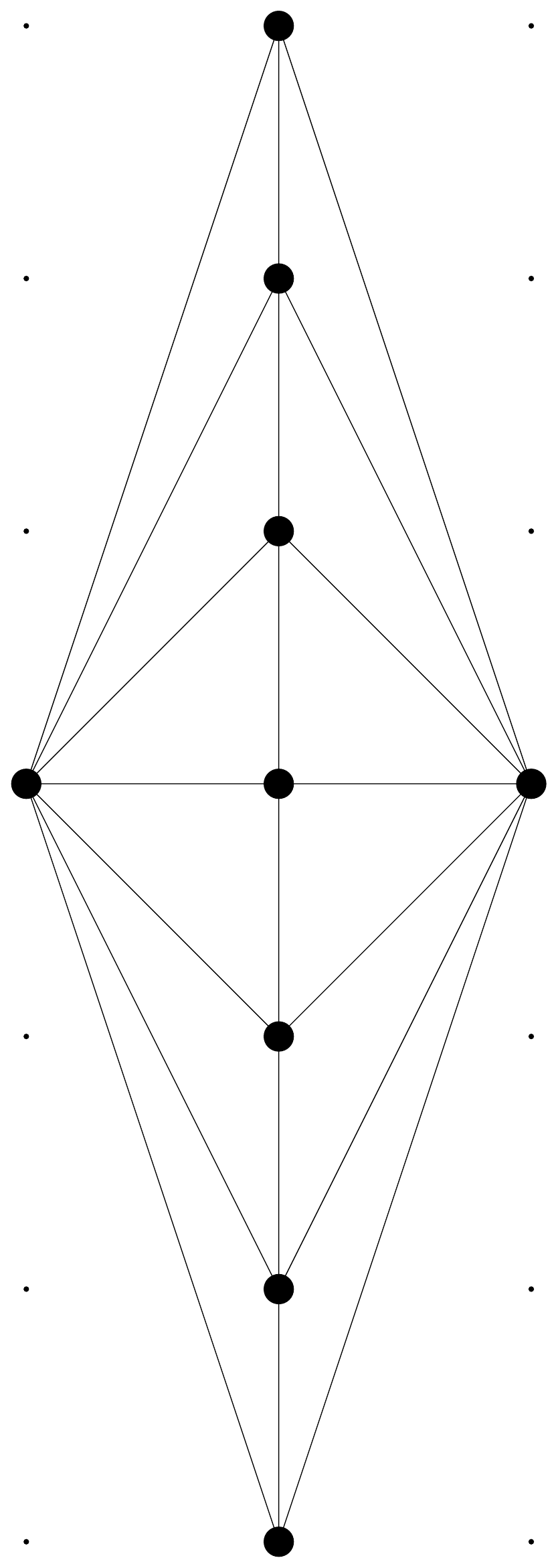}\\[50pt]
\includegraphics[width=.3\textwidth]{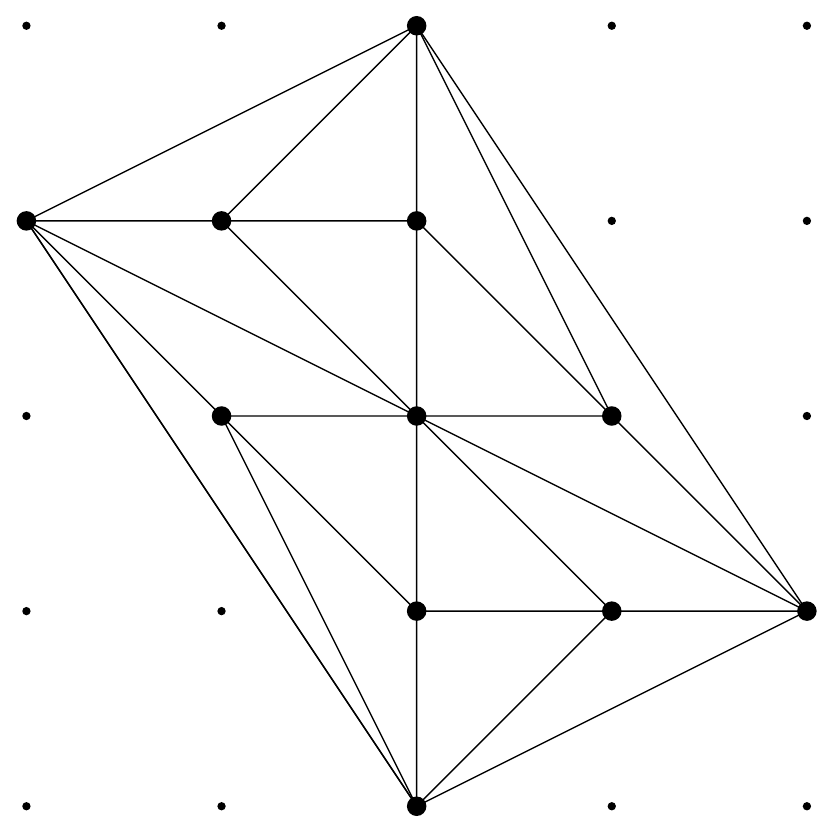}
\hskip75pt
\includegraphics[width=.3\textwidth]{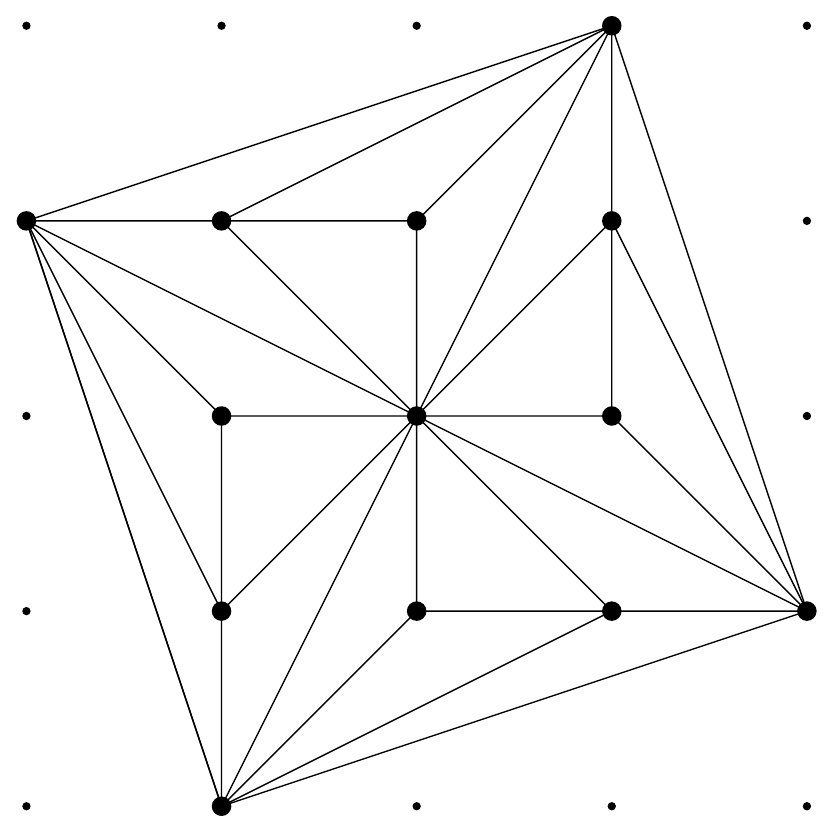}
\end{center}
}}
\parbox{6in}{\caption{\small  \label{fig:resolutions}
The fans for the K\"ahler crepant resolutions of the $\IZ_{2M}$
hyperconifold singularities, where $M = 2,3,4,5$.
}}
\end{center}
\end{figure}

We can perform the same analysis for each $\IZ_{2M}$-hyperconifold,
obtaining the fans in \fref{fig:resolutions}.  The reader may find it
amusing to follow the steps in each case, and verify the resulting fans.
At present I can provide no argument that varieties containing the $\IZ_3$-
and $\IZ_5$-hyperconifolds also admit K\"ahler crepant resolutions,
but one is naturally drawn to conjecture that this is the case.  The following
comments would then apply to these cases too.

It is easy to obtain certain topological data about these resolutions.  From
the toric diagrams we see that in each case the exceptional
set $E$ is simply connected, which is the case for any toric variety whose
fan contains a top-dimensional cone.  Therefore the resolution
of $X_0$ is simply-connected, even though the smooth Calabi-Yau $X$,
of which $X_0$ is a deformation, had fundamental group $\IZ_N$.  This
contrasts with the case of a conifold transition, where the fundamental group
does not change.

We can also simply
read off the diagram that the exceptional set of the resolution of the
$\IZ_N$-hyperconifold has Euler characteristic $\ch(E)=2N$, since $\ch$
is just the number of top-dimensional cones in the fan.  We can therefore
calculate $\ch(\widehat X)$ quite easily.  Topologically,
$\widetilde X_0$ is obtained from $\widetilde X$ by shrinking an
$S^3$ to a point $P_0$, so
$\ch(\widetilde X_0) = \ch(\widetilde X)+1$.  We delete $P_0$,
quotient by $\IZ_N$, then glue in $E$, so
\begin{equation}
\ch(\widehat X) = \ch(\widetilde X)/N + \ch(E) = \ch(X) + 2N
\end{equation}
Finally, the resolution of the $\IZ_N$-hyperconifold introduces $N-1$
new divisor classes, so we can actually calculate all the Betti numbers
of $\widehat X$ in terms of those of $X$:
\begin{equation}
b_1(\widehat X) = b_5(\widehat X) = 0~,~~
b_2(\widehat X) = b_4(\widehat X) = b_2(X) + N - 1~,~~
b_3(\widehat X) = b_3(X) - 2
\end{equation}
%

\section{Hyperconifold transitions in string theory?}
\label{sec:stringtransitions}

A natural question to ask is whether the `hyperconifold transitions'
described in this paper can be realised in string theory, as their
cousins flops and conifold transitions can.  At this stage I will merely
make some suggestive observations.\footnote{The following is superceded
by the follow-up paper \cite{Davies:2011is}.}

Consider type IIB string theory on a Calabi-Yau manifold $X$ with
fundamental group $\IZ_N$, and vary the complex structure moduli
until we approach a singular variety $X_0$.  We have seen (at
least in the simplest cases) that topologically this looks like shrinking
a three-cycle $L(N,1)$ to a point.  Therefore just as in the conifold
case, there will be D3-brane states becoming massless
\cite{Strominger:1995cz}.  This manifests in the low-energy theory
as a hypermultiplet charged under a $U(1)$ gauge group coming from
the R-R sector, and although it becomes massless there is still a
D-term potential preventing its scalars from developing a VEV.
However these D-brane states are not necessarily the only ones
becoming massless at the hyperconifold point -- there are $N-1$
twisted sectors coming from strings wrapping non-trivial loops on
$L(N,1)$, and these strings attain zero length at the singular point.
It is therefore conceivable that these twisted sectors give rise to a
new branch of the low-energy moduli space, and that moving onto this
branch corresponds to resolving the singularity of the internal space.
Since there are $N-1$ new divisors/K\"ahler parameters on the
resolution, there must be $N-1$ new flat directions.

The conjecture then is that in the low energy field theory at the
hyperconifold point, there is a new $(N-1)$-dimensional branch of
moduli space coming from the twisted sectors in the string theory.
The new flat directions correspond to K\"ahler parameters on the
resolution of the singular variety, and giving them VEVs resolves the
singularity.  It would be interesting to try to verify this picture.

\subsection*{Acknowledgements}

I am very grateful to Bal\'azs Szendr\H{o}i for guidance on constructing
resolutions.  I would also like to thank Xenia de la Ossa and Andre Lukas
for reading and commenting on the manuscript, and Volker Braun, Philip
Candelas, Yang-Hui He and James Sparks for helpful discussions.  This
research was supported by the Sir Arthur Sims Travelling Scholarship Fund
and the University College Old Members'-Oxford Australia Scholarship Fund.

\newpage

\bibliographystyle{utphys}
\bibliography{references}

\end{document}